\documentclass[aps,pra,twocolumn,showpacs,amsmath,amssymb,preprintnumbers,superscriptaddress,10pt]{revtex4-1}

\usepackage{bm}
\usepackage{graphicx}
\graphicspath{{figures/}}
\usepackage{SIunits}
\usepackage{hyphenat}
\usepackage{physics}
\usepackage{scrextend}
\usepackage{color}
\usepackage{amsmath}
\usepackage{lipsum}
\usepackage{SIunits}
\usepackage[dvipsnames]{xcolor}
\usepackage{mathtools}
\usepackage{comment}
\usepackage{dsfont}
\usepackage[capitalise]{cleveref}
\crefname{section}{Sec.}{Secs.}
\Crefname{section}{Section}{Sections}

\newcommand{\set}[1]{\{#1\}}

\newcommand{\ie}{\textit{i.e.}}

\begin{document}

\title{Practical Quantum Key Distribution Secure Against Side-Channels}

\author{\'Alvaro Navarrete}
\email{anavarrete@com.uvigo.es}
\affiliation{EI Telecomunicaci\'on, Department of Signal Theory and Communications, University of Vigo, Vigo E-36310, Spain}

\author{Margarida Pereira}
\affiliation{EI Telecomunicaci\'on, Department of Signal Theory and Communications, University of Vigo, Vigo E-36310, Spain}

\author{Marcos Curty}
\affiliation{EI Telecomunicaci\'on, Department of Signal Theory and Communications, University of Vigo, Vigo E-36310, Spain}

\author{Kiyoshi Tamaki}
\affiliation{Faculty of Engineering, University of Toyama, Gofuku 3190, Toyama 930-8555, Japan}


\begin{abstract}
There is a big gap between theory and practice in quantum key distribution (QKD) because real devices do not satisfy the assumptions required by the security proofs. Here, we close this gap by introducing a simple and practical measurement-device-independent (MDI) QKD type of protocol, based on the transmission of coherent light, for which we prove its security against any possible device imperfection and/or side-channel at the transmitters' side. Besides using a much simpler experimental set-up and source characterization with only one single parameter, we show that the performance of the protocol is comparable to other MDI-QKD type of protocols which disregard the effect of several side-channels.
\end{abstract}

\maketitle

\textit{Introduction}.---Recent years have witnessed a tremendous progress in the field of quantum key distribution (QKD)~\cite{lo2014,xu2020,pirandola2019advances}, which includes the realization of long-distance fiber-based implementations~\cite{yin2016,boaron2018secure,chen2020}, satellite links~\cite{liao2017satellite,takenaka2017satellite,yin2020}, and the deployment of QKD networks~\cite{peev2009,sasaki2011,dynes2019}. Despite these groundbreaking results, however, the security of QKD implementations has not been fully established yet, due to the difficulty of real devices to satisfy the assumptions required by the security proofs. 

To bridge this pressing gap between theory and practice in QKD, various approaches have been proposed~\cite{mayers1998,acin2007,vazirani2014,lo2012}, being measurement-device-independent (MDI) QKD~\cite{lo2012} probably the most promising one, as it can remove all assumptions about the measurement unit, arguably the Achilles' heel of QKD~\cite{lydersen2010a,jain2016}. Moreover, very recently, it has been shown that a variant of MDI-QKD, the so-called twin-field QKD~\cite{lucamarini2018overcoming,wang2018twin,curty2018simple, ma2018phase,cui2019twin,tamaki2018information,lin2018simple}, can beat the private capacity of a point-to-point QKD link~\cite{takeoka2014fundamental,pirandola2017fundamental}, thus offering unprecedented high key rates and achievable distances~\cite{chen2020,zhong2019proof,minder2019experimental,liu2019experimental,wang2019beating}. 

Nonetheless, MDI-QKD still needs that certain assumptions are satisfied. Precisely, the users (called Alice and Bob) must characterize their emitted signals accurately, and then incorporate this information in the security proof. These signals typically deviate from those prescribed by the ideal protocol due to inevitable device imperfections, and/or owing to the action of the eavesdropper (Eve), who might launch, for instance, a Trojan Horse attack (THA)~\cite{gisin2006,jain2014,lucamarini2015}. If these deviations are not taken into account, they might open security loopholes, or so-called side-channels, which could be exploited by Eve. State preparation flaws (SPFs) can be efficiently incorporated into the security analysis by means of the loss-tolerant protocol~\cite{tamaki2014,boaron2018secure,tang2016experimental,pereira2019flawed}. Also, discrete phase-randomization has been addressed in~\cite{cao2015}. Moreover, techniques to investigate the problem of information leakage about Alice and Bob's internal settings (due to, say, a THA) have been introduced in~\cite{tamaki2016,wang2018,pereira2019flawed}. More recently, methods to analyze the effect of classical pulse correlations in high-speed QKD have been presented in~\cite{yoshino2018,pereira2019correlated}. While all these works are remarkable, so far no security proof has considered all possible side-channels created by device imperfections in a practical QKD implementation. 

In this Letter, we close the gap between theory and practice in QKD by introducing a simple and practical MDI-QKD type of protocol for which we prove its security against {\it any} possible device imperfection and/or side-channel. Furthermore, besides using a much simpler experimental set-up and source characterization with only one single parameter, we show that the performance of the protocol is comparable to other MDI-QKD type of protocols which disregard the effect of several side-channels.

\begin{figure}
	\begin{center}
		\includegraphics[width=\columnwidth]{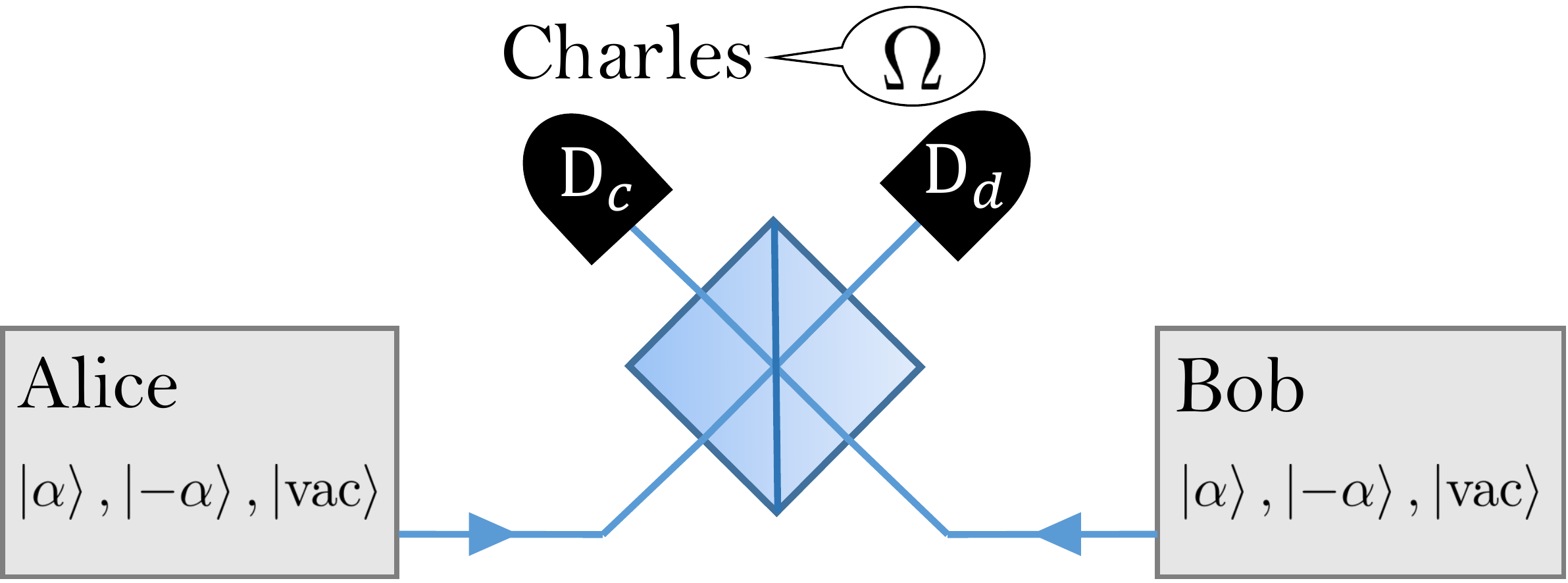}
	\end{center}
	\caption{Graphical illustration of the protocol. In each round, each of Alice and Bob randomly selects one state from the set $\set{\ket{\alpha},\ket{-\alpha},\ket{\text{vac}}}$ and sends it to Charles, who interferes the incoming signals in a 50:50 beamsplitter followed by two threshold detectors, $\text{D}_c$ and $\text{D}_d$.}\label{fig:PE_mdiQKD_scheme}
\end{figure}

\textit{Protocol description}.---For simplicity, in the protocol description we assume the ideal scenario where there are no side-channels and all the prepared states are perfect. The presence of side-channels or SPFs is discussed afterwards. That is, the description below represents an idealized scenario, and, in practice, Alice and Bob do not necessarily have to generate the states assumed here. Moreover, we consider the symmetric situation where the set of transmitted states and their a priori probabilities are equal for Alice and Bob. Also, we assume that the untrusted node Charles is located in the middle between them. We remark, however, that the generalization to the asymmetric scenario is straightforward~\cite{xu2013,liu2019experimental_demonstration,grasselli2019asymmetric,wang2019asymmetric,wang2020simple,zhong2020proof}. The setup is shown in Fig.~\ref{fig:PE_mdiQKD_scheme}.

\noindent\rule{\columnwidth}{0.3pt}
\begin{enumerate}
	\item Alice (Bob) sends a coherent state $\ket{\nu}_a$ ($\ket{\omega}_b$) to the untrusted node Charles with probability $p_{\nu}$ ($p_{\omega}$), where $\nu,\omega\in \mathcal{T}:=\set{\alpha,-\alpha,\text{vac}}$. The key states $\ket{\alpha}$ and $\ket{-\alpha}$ are associated with the bit values $0$ and $1$, respectively, and the vacuum state $\ket{\text{vac}}$ is used for parameter estimation.
	\item If Charles is honest, he interferes the incoming pulses in a 50:50 beamsplitter followed by two threshold detectors, $\text{D}_c$ and $\text{D}_d$, which are associated with constructive and destructive interference, respectively. If his measurement succeeds, which means that only one of his detectors clicks, Charles announces the measurement outcome $\Omega\in\set{\Omega_c,\Omega_d}$, where $\Omega_c$ ($\Omega_d$) corresponds to a click event only in detector $\text{D}_c$ ($\text{D}_d$). Otherwise, he announces the failure event. Besides, if $\Omega_d$ is announced, Bob flips his bit value.
	\item The previous two steps are repeated $N$ times. Next, Alice and Bob reveal their state choices for all the rounds in which at least one of them sent the vacuum state. The bits associated with the remaining rounds declared as successful by Charles constitute their sifted key.
	\item Alice and Bob announce part of their sifted key and they estimate both the bit and the phase error rates. Finally, they perform error correction and privacy amplification to obtain, with high probability, a secret key.
\end{enumerate}
\noindent\rule{\columnwidth}{0.3pt}

\textit{Side-Channels}.---Being a MDI-QKD~\cite{lo2012} type of protocol, we have that the scheme above is immune against \textit{all} detection side-channel attacks, so below we focus only on the potential side-channels at the transmitters. We begin by explaining how we describe the emitted states, and then we move on to the security proof.

For each particular round of the protocol, if Alice and Bob select, say, the settings $\nu$ and $\omega$, respectively, the joint state of their transmitted systems $a$ and $b$, and Eve's system $E$, can always be written as
\begin{equation}\label{eq:Psiij}
	\ket{\Psi_{\nu,\omega}}_{T}=\sqrt{1-\epsilon_{\nu,\omega}}\ket{\phi_{\nu,\omega}}_{T}+\sqrt{\epsilon_{\nu,\omega}}\ket{\phi^{\perp}_{\nu,\omega}}_{T},
\end{equation}
where $T:=abE$, $\epsilon_{\nu,\omega}\in[0,1]$,  $\ket{\phi_{\nu,\omega}}_{T}:=\ket{\nu}_a\ket{\omega}_b\ket{\tau}_E$ with $\ket{\tau}_E$ being a state which does not contain any information about Alice and Bob's setting choices for the current round, and $\ket{\phi^{\perp}_{\nu,\omega}}_{T}$ is a state orthogonal to $\ket{\phi_{\nu,\omega}}_{T}$. Importantly, as we show below, Eq.~(\ref{eq:Psiij}) represents the most general description of the transmitted states, which means that any potential SPF or information leakage about the internal settings of Alice and Bob can be characterized with that equation. This includes active information leakage due to, say, a THA~\cite{gisin2006,vakhitov2001,lucamarini2015,tamaki2016,wang2018}, passive information leakage due to device imperfections, or both of them simultaneously. Also, it includes classical pulse correlations, since they can be treated as passive information leakage~\cite{pereira2019correlated}, as well as coherent attacks. To see this latter fact, one can consider the purification of all systems held by Alice, Bob and Eve during the protocol. Moreover, we allow all systems held by Eve to jointly interact with all the optical pulses emitted by Alice and Bob. Also, we introduce some ancilla systems $A\bar{A}$ and $B\bar{B}$ for Alice and Bob, respectively, that contain all their setting information in an entanglement-based picture of the protocol. Here we use the notation $x$ ($\bar{x}$) to encapsulate the systems belonging to the particular round (all rounds except the particular round) that is being considered. In doing so, we have that the entire global system comprises the systems $ABab\bar{A}\bar{B}\bar{a}\bar{b} E$. Now, if Alice and Bob perform projective measurements on their ancillas $A$ and $B$ to obtain their setting information for that particular round (note that some subsystems within $\bar{A}$ and $\bar{B}$ associated with the previous rounds could have been already measured), it is straightforward to show that the resulting state for that round can be written as $\ket{\Psi_{\nu,\omega}}_{T}$ by simply redefining the joint system $\bar{A}\bar{B}\bar{a}\bar{b}E$ as $E$.

To explicitly show that Eq.~(\ref{eq:Psiij}) is indeed the most general description of the transmitted states, let $\ket{\tilde{\varphi}_{\nu,\omega}}_E$ be an unnormalized state such that $\ket{\tilde{\varphi}_{\nu,\omega}}_E= \prescript{}{a}{\bra{\nu}}\prescript{}{b}{\bra{\omega}\ket{\Psi_{\nu,\omega}}_{T}}$. Note that Eq.~(\ref{eq:Psiij}) holds trivially with $\epsilon_{\nu,\omega}=1$ if $ \prescript{}{a}{\bra{\nu}}\prescript{}{b}{\bra{\omega}\ket{\Psi_{\nu,\omega}}}_{T}=0$. Then, $\ket{\Psi_{\nu, \omega}}_{T}$ can always be written in the following form
\begin{equation}\label{eq:Psiij2}
\ket{\Psi_{\nu, \omega}}_{T}=\ket{\nu}_a\ket{\omega}_b\ket{\tilde{\varphi}_{\nu,\omega}}_E+\ket{\tilde{\chi}_{\nu,\omega}}_{T},
\end{equation}
being $\ket{\tilde{\chi}_{\nu,\omega}}_{T}$ another unnormalized state such that $\prescript{}{a}{\bra{\nu}}\prescript{}{b}{\bra{\omega}\ket{\tilde{\chi}_{\nu,\omega}}}_{T}=0$. Similarly, for some $\epsilon_{\nu,\omega}\in[0,1]$, the unnormalized state $\ket{\tilde{\varphi}_{\nu,\omega}}_E$ can always be written as
\begin{equation}\label{eq:varphi}
\ket{\tilde{\varphi}_{\nu,\omega}}_E=\sqrt{1-\epsilon_{\nu,\omega}}\ket*{\tau}_E+\ket{\tilde{\tau}^{\perp}_{\nu,\omega}}_E,
\end{equation}
where $\ket{\tau}_E$ is a normalized state which does not depend on the internal settings of the transmitters and $\ket{\tilde{\tau}^{\perp}_{\nu,\omega}}_E$ is an unnormalized state orthogonal to $\ket{\tau}_E$. Finally, by combining Eqs.~(\ref{eq:Psiij2}) and~(\ref{eq:varphi}), one directly recovers Eq.~(\ref{eq:Psiij}) with $\sqrt{\epsilon_{\nu,\omega}}\ket{\phi^{\perp}_{\nu,\omega}}_T=\ket{\tilde{\chi}_{\nu,\omega}}_{T}+\ket{\nu}_a\ket{\omega}_b\ket{\tilde{\tau}^{\perp}_{\nu,\omega}}_E$. 

Let us conclude this part by further illustrating the meaning of Eq.~(\ref{eq:Psiij}) with a simple example. For instance, suppose a THA where $\ket{\Psi_{\nu, \omega}}_{T}=\ket{\nu}_a\ket{\omega}_b\ket{\Lambda_{\nu,\omega}}_E$, being $\ket{\Lambda_{\nu,\omega}}_E$ the state of the back-reflected light which carries information about the transmitters' settings. The state $\ket{\Lambda_{\nu,\omega}}_E$ can always be written as a superposition of the vacuum state and a state $\ket{\Lambda_{\nu,\omega}'}_E$ that contains no vacuum component, \textit{i.e.}, $\ket{\Psi_{\nu,\omega}}_{T} = \lambda\ket{\nu}_a\ket{\omega}_b\ket{\text{vac}}_E + \sqrt{1-\lambda^2}\ket{\nu}_a\ket{\omega}_b\ket{\Lambda_{\nu,\omega}'}_E$. This is so due to inevitable losses at the transmitters (\textit{e.g.}, produced by material absorption
or due to the presence of optical isolators), which guarantee $\lambda>0$. This latter equation is equivalent to Eq.~(\ref{eq:Psiij}) for $\ket{\tau}_E=\ket{\text{vac}}_E$.

\textit{Security proof}.---To prove the security of the protocol above, we shall assume that Alice and Bob know an upper bound on $\epsilon_{\nu,\omega}$ for each round, but no characterization is needed for the side-channel states $\ket*{\phi_{\nu,\omega}^{\perp}}_{T}$ in Eq.~(\ref{eq:Psiij}). We remark, however, that any available information about the states $\ket{\phi_{\nu,\omega}^{\perp}}_{T}$ could be readily incorporated in the security proof described below. Also, we emphasize that the security proof is valid even if the states that Alice and Bob generate in the ideal scenario (\ie, without side-channels) are not $\ket{\nu}_a$ and $\ket{\omega}_b$, or they are mixed states, due, for instance, to SPFs. In other words,  $\ket{\nu}_a$ and $\ket{\omega}_b$ are adopted in Eq.~(\ref{eq:Psiij}) just as a reference for the state characterization in the experiment.

To calculate a lower bound on the secret key rate of the protocol, we first need to estimate the  phase error rate $e_{\text{ph}}$, which is a key parameter in the complementarity argument~\cite{koashi2009}. For this, note that, from Eve's perspective, the actual scenario where both Alice and Bob send Charles key states is equivalently described by a fictitious scenario where, instead, they first prepare the entangled state 
\begin{equation}\label{eq:VirtualGlobal}
\begin{split}
\ket{\Psi^{\text{vir}}}_{ABT}=\frac{1}{2}\sum_{j,s=0,1}\ket{j_z,s_z}_{AB}\ket{\Psi_{(-1)^j\alpha,(-1)^s\alpha}}_{T},
\end{split}
\end{equation}
with $\set{\ket{0_z},\ket{1_z}}$ being the computational basis for the ancilla systems $A$ and $B$, and subsequently they send the system $T$ to Charles. This equivalence holds because measurements on the ancilla systems $A$ and $B$ commute with those on the system $T$. Here, and in what follows, we shall consider that $j,s\in\set{0,1}$ when referring to the virtual states. Now, we can imagine a fictitious virtual scenario where Alice and Bob measure their ancillas $A$ and $B$ in the complementary basis $\set{\ket{0_x},\ket{1_x}}$, being $\ket{j_x}=1/\sqrt{2}\left[\ket{0_z}+(-1)^j\ket{1_z}\right]$. In this virtual scenario, the unnormalized reduced density operators of the transmitted states are given by 
\begin{equation}
\bar{\sigma}_{j,s}^{\text{vir}}=\Tr_{AB}\left[\dyad{j_x,s_x}{j_x,s_x}_{AB}\otimes\mathds{1}_{T}\dyad{\Psi^{\text{vir}}}{\Psi^{\text{vir}}}_{ABT}\right],
\end{equation}
where $\mathds{1}_{T}$ is the identity operator acting on $T$. We call the states $\bar{\sigma}_{j,s}^{\text{vir}}$ the unnormalized virtual states, and we write their normalized form as $\sigma_{j,s}^{\text{vir}}\equiv\ketbra{\Psi_{j,s}^{\text{vir}}}_{T}$. 

The phase error rate is then defined as the bit error rate of the virtual scenario. In the protocol above, a phase error occurs when Alice and Bob measure either $\ket{0_x,0_x}_{AB}$ or $\ket{1_x,1_x}_{AB}$  and Charles announces a successful event (see \cref{appendix:channelmodel}). This means that
\begin{equation}\label{eq:phaseError}
e_{\text{ph}}=\frac{p_{0,0}^{\rm vir}Y_{0,0}^{\text{vir}}+p_{1,1}^{\rm vir}Y_{1,1}^{\text{vir}}}{\sum_{j,s}p_{j,s}^{\rm vir}Y_{j,s}^{\text{vir}}},
\end{equation}
where $Y_{j,s}^{\text{vir}}$ is the conditional probability of a successful announcement by Charles given that Alice and Bob send $\sigma_{j,s}^{\text{vir}}$, and $p_{j,s}^{\rm vir}=\Tr\{\bar{\sigma}_{j,s}^{\text{vir}}\}$. Note that, since Alice and Bob measure their ancillas in the complementary basis, the bit flip operation performed by Bob when Charles announces a result $\Omega_d$ has no effect in the virtual scenario. The term $\sum_{j,s}p_{j,s}^{\rm vir}Y_{j,s}^{\text{vir}}=:\gamma_{\text{obs}}$ in Eq.~(\ref{eq:phaseError}) is equal to the probability that Charles announces a successful event and both Alice and Bob send a key state. This quantity is directly observed in the actual experiment. Thus, to calculate $e_{\text{ph}}$ it is enough to estimate the phase error probability $p_{0,0}^{\rm vir}Y_{0,0}^{\text{vir}}+p_{1,1}^{\rm vir}Y_{1,1}^{\text{vir}}=:\Gamma$.

For this, we use the reference technique recently introduced in~\cite{pereira2019correlated}. Specifically, we first define, for each user, a set of qubit states $\set{\ket{\Phi_{\alpha}},\ket{\Phi_{-\alpha}},\ket{\Phi_{\text{vac}}}}$ called the reference states. We have freedom to select the reference states, however, for the security proof to go through, a lower bound on $\abs{\braket{\Phi_{\nu, \omega}}{\Psi_{\nu, \omega}}}$ for each possible combination of $\nu$ and $\omega$ is needed, being $\ket{\Phi_{\nu,\omega}}_T:=\ket{\Phi_{\nu}}_a\otimes\ket{\Phi_{\omega}}_b\otimes\ket{\tau}_E$. That is, the joint reference states $\ket{\Phi_{\nu,\omega}}_T$ should be chosen similar to the original transmitted states $\ket{\Psi_{\nu,\omega}}_T$, which in practice is equivalent to say that they should be similar to the states $\ket{\phi_{\nu,\omega}}_T$. In what follows, we will omit the mode subscripts for readability whenever is clear.

A natural choice for the set of reference states is given by $\set{\ket{\alpha},\ket{-\alpha},\ket{\text{vac}'}}$, where the state $\ket{\text{vac}'}$ is the projection of $\ket{\text{vac}}$ onto the qubit space spanned by $\set{\ket{\alpha},\ket{-\alpha}}$. For this, let the orthonormal basis $\set{\ket{0_{\text{o}}},\ket{1_{\text{o}}},\ket{2_{\text{o}}}}$ satisfy
\begin{eqnarray}
\begin{split}
\ket{\alpha}&=\ket{0_{\text{o}}},\\
\ket{-\alpha}&=\braket{\alpha}{-\alpha}\ket{0_{\text{o}}}+\sqrt{1-\abs{\braket{\alpha}{-\alpha}}^2}\ket{1_{\text{o}}},\\
\ket{\text{vac}}&=\braket{\alpha}{\text{vac}}\ket{0_{\text{o}}}+c_1\ket{1_{\text{o}}}+c_2\ket{2_{\text{o}}},
\end{split}
\end{eqnarray}
where the coefficients $c_1$ and $c_2$ fulfil $\braket{-\alpha}{\text{vac}}=\braket{-\alpha}{\alpha}\braket{\alpha}{\text{vac}}+c_1\sqrt{1-\abs{\braket{\alpha}{-\alpha}}^2}$ and
$\abs{\braket{\alpha}{\text{vac}}}^2+\abs{c_1}^2+c_2^2=1$, and where, without loss of generality, we assume that $c_2$ is real. This means, in particular, that $\ket{\text{vac}'}=1/\sqrt{\xi}\left[\braket{\alpha}{\text{vac}}\ket{0_{\text{o}}}+c_1\ket{1_{\text{o}}}\right]$, with $\xi=\abs{\braket{\alpha}{\text{vac}}}^2+\abs{c_1}^2$.

From the definitions of $\ket{\Psi^{\text{vir}}}$, $\sigma_{j,s}^{\text{vir}},p_{j,s}^{\text{vir}}$ and $Y_{j,s}^{\text{vir}}$, one can define analogous states and probabilities $\ket{\Phi^{\text{vir}}}$, $\sigma_{j,s}^{\text{vir}|\text{ref}}$, $p_{j,s}^{\text{vir}|\text{ref}}$ and $Y_{j,s}^{\text{vir}|\text{ref}}$ for the reference states above by simply substituting the actual states $\ket{\Psi_{\nu,\omega}}$ with the reference states $\ket{\Phi_{\nu,\omega}}$ where needed in their definitions~\cite{pereira2019correlated}. In particular, the yields $Y_{j,s}^{\text{vir}|\text{ref}}$ are defined as
\begin{equation}\label{eq:Yjs}
Y_{j,s}^{\text{vir}|\text{ref}}=\Tr\left[\hat{\mathcal{D}}\sigma_{j,s}^{\text{vir}|\text{ref}}\right],
\end{equation}
where $\hat{\mathcal{D}}$ is the POVM element associated with Charles' successful announcement. Now, to estimate $\Gamma$, one can define an analogous quantity for the reference states, namely $\Gamma_{\text{ref}}:= p_{0,0}^{\rm vir|\text{ref}}Y_{0,0}^{\text{vir}|\text{ref}}+p_{1,1}^{\rm vir|\text{ref}}Y_{1,1}^{\text{vir}|\text{ref}}$, and then quantify the maximum possible deviation in the measurement statistics between the reference and the actual scenario. For this, we conveniently define the operator $\hat{\mathcal{D}}_{\text{ph}}=(\ketbra{0_x,0_x}+\ketbra{1_x,1_x})\otimes\hat{\mathcal{D}}$ and then we use the fact that, for any operator $0 \preceq\hat{\mathcal{O}}\preceq \mathds{1}$, and normalized pure states $\ket{A}$ and $\ket{R}$, the following inequality is satisfied~\cite{pereira2019correlated}
\begin{eqnarray}\label{eq:IneqRA}
\delta\leq \sqrt{Y_{A}Y_{R}}+\sqrt{(1-Y_{A})(1-Y_{R})},
\end{eqnarray}
where $\delta=\abs{\braket{\text{A}}{\text{R}}}$, $Y_{A}=\bra{\text{A}}\hat{\mathcal{O}}\ket{\text{A}}$ and $Y_{R}=\bra{\text{R}}\hat{\mathcal{O}}\ket{\text{R}}$. From Eq.~(\ref{eq:IneqRA}) one can derive the functions
\begin{eqnarray}
G_{+}(Y_R,\delta)&=
\left\{ \begin{array}{ll}
g_{+}(Y_R,\delta),\quad &Y_R<\delta^2 \\
1,\quad &\text{otherwise}
\end{array} \right.\\
G_{-}(Y_R,\delta)&=
\left\{ \begin{array}{ll}
g_{-}(Y_R,\delta),\quad &Y_R>1-\delta^2 \\
0,\quad &\text{otherwise}
\end{array} \right.
\end{eqnarray}
such that $G_{-}(Y_R,\delta)\leq Y_{A}\leq G_{+}(Y_R,\delta)$, where
$g_{\pm}(Y,\delta)=Y+(1-\delta^2)(1-2Y)\pm2\delta\sqrt{(1-\delta^2)Y(1-Y)}$.
Furthermore, given $Y^{\text{U}}\geq Y$ and $0\leq\delta^L\leq \delta$, it holds that $G_+(Y^{\text{U}},\delta^{\text{L}})\geq G_+(Y,\delta)$. Then, by noticing that $\Gamma=\bra{\Psi^{\text{vir}}}\hat{\mathcal{D}}_{\text{ph}}\ket{\Psi^{\text{vir}}}$ and $\Gamma_{\text{ref}}=\bra{\Phi^{\text{vir}}}\hat{\mathcal{D}}_{\text{ph}}\ket{\Phi^{\text{vir}}}$, an upper bound on $\Gamma$ can be simply obtained as
\begin{eqnarray}\label{eq:final}
	\Gamma&\leq& G_+(\Gamma_{\text{ref}},\delta_{\text{vir}})\nonumber\\
	&\leq& G_+(\Gamma_{\text{ref}}^{\text{U}},\delta_ {\text{vir}}^{\text{L}})=:\Gamma^{\text{U}},
\end{eqnarray}
where $\Gamma_{\text{ref}}^{\text{U}}$ is an upper bound on $\Gamma_{\text{ref}}$ (see \cref{appendix:Bounds} for a particular expression) and $\delta_{\text{vir}}^{\text{L}}=1/4\sum_{j,s=0,1}\sqrt{1-\epsilon_{(-1)^j\alpha,(-1)^s\alpha}}$ is a lower bound on $\delta_{\text{vir}}:=\abs{\braket{\Phi^\text{vir}}{\Psi^\text{vir}}}$.

Importantly, it can be shown that $\Gamma^{\text{U}}$ can be written as a concave function of the observed statistics $Y_{\nu,\omega}:=\bra{\Psi_{\nu,\omega}}\hat{\mathcal{D}}\ket{\Psi_{\nu,\omega}}$ and, therefore, the security of the protocol can be easily extended against coherent attacks. We refer the reader to \cref{appendix:coherent} for further details.

Finally, given an upper bound $e_{\text{ph}}^\text{U}=\Gamma^{\text{U}}/\gamma_{\text{obs}}$ on $e_{\text{ph}}$, the asymptotic secret key rate can be written as
\begin{equation}
	R\geq Q[1-h(e_{\text{ph}}^\text{U})-f_e h(e_{\text{bit}})],
\end{equation}
where $e_{\text{bit}}$ is the bit error rate, $f_e$ is the error correction efficiency and $Q$ is the probability that both Alice and Bob select a key state and Charles announces a successful event.

\begin{figure}
	\begin{center}
		\includegraphics[width=\columnwidth]{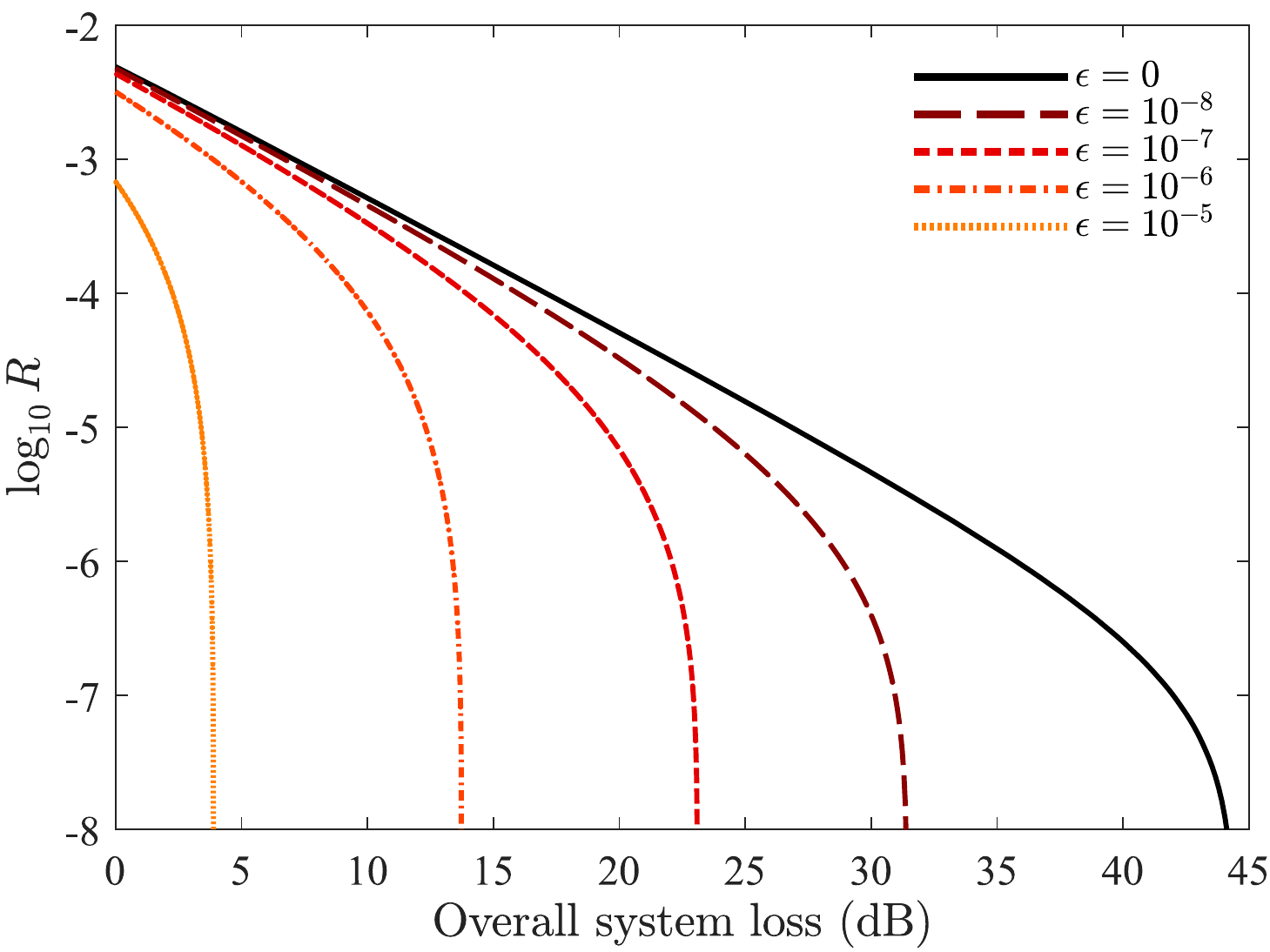}
	\end{center}
	\caption{Secret key rate $R$ as a function of the overall system loss (in $\deci\bel$) between Alice and Bob for different values of the parameter $\epsilon$. For simplicity, we consider the symmetric scenario where Charles is located in the middle between Alice and Bob. The value of $\alpha$ has been optimized for each system loss value.}\label{fig:Fig_Results}
\end{figure}

\textit{Evaluation}.---Fig.~\ref{fig:Fig_Results} shows the secret key rate of the protocol in the presence of side-channels. For simplicity, here we set $\epsilon_{\nu,\omega}=\epsilon$ for all $\nu,\omega\in \mathcal{T}$, and we optimize the parameter $\alpha$ for each value of the overall system loss. In our simulations, we model system loss with a beamsplitter and, also, for simplicity, we disregard any misalignment effect in the channel. In addition, we set the dark-count probability of Charles' detectors to $p_d=10^{-8}$ to match some recent experiments~\cite{minder2019experimental}. Further details about the channel model and the optimal values for $\alpha$ can be found in the Appendices~\ref{appendix:channelmodel} and~\ref{appendix:alpha}.

As expected, the performance of the protocol decreases when $\epsilon$ increases. Also, Fig.~\ref{fig:Fig_Results} shows that, for the channel model considered, a positive secret key rate is possible even when $\epsilon=10^{-5}$. Note that $\epsilon$ characterizes, for each state $\ket{\Psi_{\nu,\omega}}$, the information leakage of both users. For instance, when $\epsilon=10^{-6}$ our simulation results suggest that Alice and Bob could generate a secret key over about 14 $\deci\bel$ of overall system loss, which corresponds to a transmission distance of about 50 $\kilo\metre$ when considering threshold detectors with 44\% of detection efficiency~\cite{minder2019experimental} and standard optical fibres with loss coefficient $0.2$ $\deci\bel/\kilo\metre$. If $\epsilon$ is sufficiently small, the tolerance of the protocol against the system loss becomes comparable to some MDI-QKD protocols which assume that the transmitted states are characterized precisely~\cite{xu2014a,tang2016experimental}. Also, its key rate is greater than that of the standard MDI-QKD scheme assuming leaky sources~\cite{wang2020measurement}, even though this latter work assumes that there are no SPFs, Alice and Bob apply perfect phase randomization and this phase information is not leaked to Eve. Moreover, we remark that this is achieved without requiring the use of the decoy-state technique~\cite{hwang2003,lo2005,wang2005a} nor the use of phase randomized coherent pulses, which could open additional side-channels that Eve might exploit~\cite{tamaki2016,wang2018,huang2019laser,sun2012,sun2015}. Furthermore, we note that Charles' station is also simpler, as it only requires two detectors (rather than four) to distinguish two Bell states~\cite{tamaki2012phase,lucamarini2018overcoming}.

The simulations in Fig.~\ref{fig:Fig_Results} assume that Alice and Bob can emit perfect vacuum signals when $\epsilon=0$ and we consider side-channels attached to the perfect vacuum states only for simplicity. In practice, however, due to the finite extinction ratio of intensity modulators, it might be difficult for them to generate a perfect vacuum state. Importantly, we note that very similar results as those illustrated in Fig.~\ref{fig:Fig_Results} can be obtained if Alice and Bob replace the vacuum signals with sufficiently weak coherent states (see \cref{appendix:vacuum} for further details).

\textit{Conclusions}.---We have presented a simple and practical MDI-QKD type of protocol which can accommodate any transmitter imperfections in the security proof, thus closing the gap between theoretical and implementation security in QKD. Moreover, it can offer a performance comparable to other MDI-QKD type of solutions by using a simpler set-up which only requires the emission of coherent light pulses and two threshold detectors at the intermediate node.

\textit{Aknowledgements}.---We thank Guillermo Curr\'as-Lorenzo for valuable discussions. This work was supported by the European Union's Horizon 2020 research and innovation programme under the Marie Sklodowska-Curie grant agreement number 675662 (Project QCALL). M.C. also acknowledges support from the Spanish Ministry of Economy and Competitiveness (MINECO), and the Fondo Europeo de Desarrollo Regional (FEDER) through the grant TEC2017-88243-R. A.N. acknowledges support from a FPU scholarship from the Spanish Ministry of Education. K.T. acknowledges support from JSPS KAKENHI Grant Numbers JP18H05237 18H05237 and JST-CREST JPMJCR 1671.

\def\bibsection{\medskip\begin{center}\rule{0.5\columnwidth}{.8pt}\end{center}\medskip}
\bibliography{library}

\appendix
\section{Channel model}\label{appendix:channelmodel}
Here we present the expected values for the quantities $Y_{\nu,\omega}$ used to estimate the phase error rate. For this, we model the loss from Alice (Bob) to Charles with a beamsplitter of transmittance $\sqrt{\eta}$, \textit{i.e.}, the overall system loss is equal to $10\log(1/\eta)$. We further assume, for simplicity, that Charles' detectors have the same dark-count probability $p_d$, and we disregard the effect of phase misalignment introduced by the channel. In this scenario, it can be shown that the conditional probability that Charles observes a click in the detector $\text{D}_c$ but not in the detector $\text{D}_d$ given that Alice and Bob send him the states $\ket{\nu}$ and $\ket{\omega}$, respectively, is given by
\begin{eqnarray}\label{eq:Ycm}
Y_{\nu,\omega}&=&(1-p_d)^2e^{-\sqrt{\eta}\left(\frac{\abs{\nu}^2+\abs{\omega}^2}{2}-\abs{\nu}\abs{\omega}\cos(\phi_A-\phi_B)\right)}\nonumber\\
&&\times\left(1-e^{-\sqrt{\eta}\left(\frac{\abs{\nu}^2+\abs{\omega}^2}{2}+\abs{\nu}\abs{\omega}\cos(\phi_A-\phi_B)\right)}\right)\nonumber\\
&&+p_d(1-p_d),
\end{eqnarray}
where $\phi_A=\arg(\nu)$ and $\phi_B=\arg(\omega)$. The same probability given by Eq.~(\ref{eq:Ycm}) is valid for the case where Charles observes destructive interference if one takes into account the bit flip at Bob's side (which is equivalent to flipping the phase of $\omega$). On the other hand, the bit error rate is given by
\begin{equation}
e_{\text{bit}}=\frac{p_d}{2p_d+e^{2\sqrt{\eta}\alpha^2}-1}.
\end{equation}

\subsubsection{Phase error}
Here we sketch how to decide the most convenient definition of a phase error in this protocol. For this, we assume the ideal scenario without side-channels. This means that, in the entanglement-based picture, the state shared by Alice and Bob in the key rounds can be written as
\begin{eqnarray}
\ket{\Psi^{\text{vir}}}&=&\frac{1}{2}\left[\ket{0_z0_z}_{AB}\ket{\alpha,\alpha}_{ab} +\ket{0_z1_z}_{AB}\ket{\alpha,-\alpha}_{ab}\right.\nonumber\\ &&\left.+\ket{1_z0_z}_{AB}\ket{-\alpha,\alpha}_{ab} +\ket{1_z1_z}_{AB}\ket{-\alpha,-\alpha}_{ab}\right].\nonumber\\
\end{eqnarray}
The beamsplitter at Charles acts on the input modes $a$ and $b$ as $\hat{a}^{\dagger}\to\frac{1}{\sqrt{2}}[\hat{c}^{\dagger}+\hat{d}^{\dagger}]$ and $\hat{b}^{\dagger}\to\frac{1}{\sqrt{2}}[\hat{c}^{\dagger}-\hat{d}^{\dagger}]$, being $c$ and $d$ the output modes corresponding to constructive and destructive interference, respectively, and where $\hat{m}^{\dagger}$ denotes the creation operator on mode $m$. Then, in an ideal scenario with no loss, the state after the beamsspliter can be written as
\begin{eqnarray}
\ket{\Psi^{\text{vir}}}&=&\frac{1}{2}\left[\ket{0_z0_z}_{AB}\ket*{\sqrt{2}\alpha}_{c}+\ket{1_z1_z}_{AB}\ket*{-\sqrt{2}\alpha}_{c}\right]\nonumber\\
&&+\frac{1}{2}\left[\ket{0_z1_z}_{AB}\ket*{\sqrt{2}\alpha}_{d}+\ket{1_z0_z}_{AB}\ket*{-\sqrt{2}\alpha}_{d}\right].\nonumber\\
\end{eqnarray}
This means that the state associated with a click in $\text{D}_c$ and no click on $\text{D}_d$ is given by
\begin{widetext}
	\begin{eqnarray}
	\ket{\Psi_c}&=&\frac{e^{\alpha^2}}{\sqrt{1-e^{2\alpha^2}}}\sum_{\substack{n=1 \\ n \text{ odd}}}^{\infty}\left[\frac{(\sqrt{2}\alpha)^n}{n!}\ket{n}_{c}\right]\otimes\frac{1}{\sqrt{2}}\left(\ket{0_z0_z}_{AB}-\ket{1_z1_z}_{AB}\right)\nonumber\\
	&&+\frac{e^{\alpha^2}}{\sqrt{1-e^{2\alpha^2}}}\sum_{\substack{n=2 \\ n \text{ even}}}^{\infty}\left[\frac{(\sqrt{2}\alpha)^n}{n!}\ket{n}_{c}\right]\otimes\frac{1}{\sqrt{2}}\left(\ket{0_z0_z}_{AB}+\ket{1_z1_z}_{AB}\right),
	\end{eqnarray}
\end{widetext}
where $\ket{n}_{c}$ is the Fock state with $n$ photons on mode $c$. The previous state can be approximated, for $\alpha$ small, to
\begin{eqnarray}
\ket{\Psi_c}&\approx&\frac{e^{\alpha^2}\alpha}{\sqrt{1-e^{2\alpha^2}}}\left[\ket{0_z0_z}_{AB}-\ket{1_z1_z}_{AB}\right]\otimes\ket{1}_{c}\nonumber\\
&=&\frac{e^{\alpha^2}\alpha}{\sqrt{1-e^{2\alpha^2}}}\left[\ket{0_x1_x}_{AB}+\ket{1_x0_x}_{AB}\right]\otimes\ket{1}_{c}.
\end{eqnarray}
Similarly, we can obtain exactly the same result for $\text{D}_d$ if we take into account Bob's bit flip. This indicates that a phase error should be defined by Alice and Bob as observing identical outcomes (\textit{i.e.}, either $\ket{0_x0_x}$ or $\ket{1_x1_x}$) in the virtual scenario.

\section{Derivation of $\Gamma_{\text{ref}}^{\text{U}}$}\label{appendix:Bounds}
Here we show how to obtain a simple upper bound on the quantity $\Gamma_{\text{ref}}$. For this, we first relate this quantity to the probabilities $Y_{j,s}^{\text{vir}|\text{ref}}$. We do so by rewritting the virtual states $\sigma_{j,s}^{\text{vir}|\text{ref}}$ as $\sigma_{j,s}^{\text{vir}|\text{ref}}=\frac{1}{4}\sum_{i,k}S_{i,k}^{j,s|\text{vir}}\sigma_i^a\otimes\sigma_k^b$, being $\sigma_i^a$ and $\sigma_k^b$ the Pauli operators with $i,k\in\set{\mathcal{I},X,Z}$, and the terms $S_{i,k}^{j,s|\text{vir}}$ the Bloch coefficients of the virtual states $\sigma_{j,s}^{\text{vir}|\text{ref}}$. Here the Pauli operator $\sigma_Y$ is not necessary because none of the states $\sigma_{j,s}^{\text{vir}|\text{ref}}$ has complex components. Thus, Eq.~(\ref{eq:Yjs}) can be rewritten as
\begin{eqnarray}\label{eq:yields-qv}
Y_{j,s}^{\text{vir}|\text{ref}}&=&\sum_{i,k}S_{i,k}^{j,s|\text{vir}}q_{i,k},
\end{eqnarray}
where $q_{i,k}=\frac{1}{4}\Tr\left[\hat{\mathcal{D}}\sigma_i^a\otimes\sigma_k^b\right]$. With this notation, one can conveniently write the following matrix equation
\begin{equation}\label{eq:matrixEq1}
\Gamma_{\text{ref}}=(\mathbf{P}^{\rm vir})^{\text{T}}\mathbf{S}^{\rm vir}\mathbf{q},
\end{equation}
where $(\mathbf{P}^{\rm vir})^{\text{T}}=[p_{0,0}^{\rm vir|\text{ref}},p_{1,1}^{\rm vir|\text{ref}}]$, $\mathbf{S}^{\rm vir}$ is a $2\times9$ matrix containing the coefficients $S_{i,k}^{0,0|\text{vir}}$ ($S_{i,k}^{1,1|\text{vir}}$) in its first (second) row, and $\mathbf{q}$ is a column vector containing the quantities $q_{i,k}$. Moreover, and analogously to Eq.~(\ref{eq:yields-qv}), one can write
\begin{eqnarray}\label{eq:yields-qo}
Y_{\nu,\omega}^{\text{ref}}&=&\sum_{i,k}S_{i,k}^{\nu,\omega}q_{i,k},
\end{eqnarray}
where $S_{i,k}^{\nu,\omega}$ denote the Bloch coefficients of the reference states $\ket{\Phi_{\nu,\omega}}$, and the quantities $Y_{\nu,\omega}^{\text{ref}}$ are their respective yields. From Eq.~(\ref{eq:yields-qo}), we find another matrix equation involving $\mathbf{q}$. It reads
\begin{equation}\label{eq:matrixEq2}
\mathbf{Y}^{\rm ref}=\mathbf{S}\mathbf{q},
\end{equation}
where $\mathbf{Y}^{\rm ref}$ is a column vector containing the yields $Y_{\nu,\omega}^{\text{ref}}$ and 
$\mathbf{S}$ is a $9\times 9$ matrix containing the Bloch coefficients of the reference states $\ket{\Phi_{\nu,\omega}}$ in its rows. Then, by combining Eqs.~(\ref{eq:matrixEq1}) and~(\ref{eq:matrixEq2}), one obtains
\begin{eqnarray}\label{eq:w_YR}
\Gamma_{\text{ref}}=(\mathbf{P}^{\rm vir})^{\text{T}}\mathbf{S}^{\rm vir}\mathbf{S}^{-1}\mathbf{Y}^{\rm ref}=\mathbf{f}_{\rm obj}\mathbf{Y}^{\rm ref},
\end{eqnarray}
where $\mathbf{f}_{\rm obj}:=(\mathbf{P}^{\rm vir})^{\text{T}}\mathbf{S}^{\rm vir}\mathbf{S}^{-1}$ is a row vector. Note that the matrix $\mathbf{S}$ is invertible because it can be written as the tensor product of two $3\times 3$ invertible matrices.

Now, to obtain an upper bound on $\Gamma_{\text{ref}}$, we bound each term in Eq.~(\ref{eq:w_YR}) separately. Specifically, we have that
\begin{eqnarray}\label{eq:GammaU}
\Gamma_{\text{ref}}&=&\mathbf{f}_{\rm obj}\mathbf{Y}^{\rm ref}\nonumber\\
&=&\sum_{\nu,\omega}f_{\nu,\omega}Y_{\nu,\omega}^{\text{ref}}\nonumber\\
&\leq&\sum_{\nu,\omega | f_{\nu,\omega}>0}f_{\nu,\omega}G_ +(Y_{\nu,\omega},\delta_{\nu,\omega}^{\rm L})\nonumber\\
&&+\sum_{\nu,\omega | f_{\nu,\omega}<0}f_{\nu,\omega}G_ -(Y_{\nu,\omega},\delta_{\nu,\omega}^{\rm L})\nonumber\\
&=:&\Gamma_{\text{ref}}^{\text{U}},
\end{eqnarray}
where the coefficients $f_{\nu,\omega}$ are the elements of the vector $\mathbf{f}_{\rm obj}$, the observed statistics $Y_{\nu,\omega}=\bra{\Psi_{\nu,\omega}}\hat{\mathcal{D}}\ket{\Psi_{\nu,\omega}}$ and the terms $\delta_{\nu,\omega}^{\rm L}$ are lower bounds on $\delta_{\nu,\omega}=\abs{\braket{\Psi_{\nu,\omega}}{\Phi_{\nu,\omega}}}$. To obtain particular expressions for the latter bounds, we first note that $\abs*{\braket{\Psi_{\nu,\omega}}{\Phi_{\nu,\omega}}}=\abs*{\sqrt{1-\epsilon_{\nu,\omega}}\braket{\phi_{\nu,\omega}}{\Phi_{\nu,\omega}}+\sqrt{\epsilon_{\nu,\omega}}\braket{\phi_{\nu,\omega}^{\perp}}{\Phi_{\nu,\omega}}}$. Now, the reference states $\ket{\Phi_{\nu,\omega}}$ can always be written as
\begin{equation}
\ket{\Phi_{\nu,\omega}}=\varsigma_{\nu,\omega}\ket{\phi_{\nu,\omega}}+\sqrt{1-\abs*{\varsigma_{\nu,\omega}}^2}\ket*{\tilde{\phi}_{\nu,\omega}^{\perp}},
\end{equation} 
where $\varsigma_{\nu,\omega}=\braket*{\phi_{\nu,\omega}}{\Phi_{\nu,\omega}}$ and $\ket*{\tilde{\phi}_{\nu,\omega}^{\perp}}$ is some state orthogonal to $\ket{\phi_{\nu,\omega}}$. Then, $\delta_{\nu,\omega}$ can be written as
\begin{equation}
\delta_{\nu,\omega}=\abs*{\sqrt{1-\epsilon_{\nu,\omega}}\varsigma_{\nu,\omega}+\sqrt{\epsilon_{\nu,\omega}}\sqrt{1-\abs*{\varsigma_{\nu,\omega}}^2}\braket*{\phi_{\nu,\omega}^{\perp}}{\tilde{\phi}_{\nu,\omega}^{\perp}}}.
\end{equation}
In our particular case, $\varsigma_{\nu,\omega}$ depends on the parameter $\xi$ defined in the main text. Specifically,  $\varsigma_{\nu,\omega}=\xi$ when $\nu=\omega=\text{vac}$, $\varsigma_{\nu,\omega}=\sqrt{\xi}$ when either $\omega\neq\nu=\text{vac}$ or $\nu\neq\omega=\text{vac}$, and $\varsigma_{\nu,\omega}=1$ otherwise. Thus, a lower bound on $\delta_{\nu,\omega}$ is straightforwardly given by
\begin{equation}
\delta_{\nu,\omega}^{\rm L}=\sqrt{1-\epsilon_{\nu,\omega}}\varsigma_{\nu,\omega}-\sqrt{\epsilon_{\nu,\omega}}\sqrt{1-\abs*{\varsigma_{\nu,\omega}}^2}.
\end{equation}

\section{Security against coherent attacks}\label{appendix:coherent}
Here we briefly show that the analysis presented in the main text can be used to guarantee security against coherent attacks. For this, note that for a protocol with $N$ rounds, Eq.~(\ref{eq:final}) is still valid for each particular round $n=1,\dots,N$. Also, let us define $p_{\mathcal{K}}$ to be the probability that a round is selected for key generation. That is, this is the probability that in a successful round none of Alice and Bob select the vacuum states nor the round is chosen to estimate the bit error rate or the phase error rate. The probability $p_{\mathcal{K}}$ can be included as a factor on the right hand side of Eq.~(\ref{eq:final}), so we obtain an upper bound on the probability that the round $n$ is used for key generation and a phase error occurs, namely  $\Gamma^{(n)}_{\mathcal{K}}$. That is,
\begin{eqnarray}
\Gamma^{(n)}_{\mathcal{K}}=p_{\mathcal{K}}\Gamma^{\text{U}}_n=p_{\mathcal{K}} G_+\left(\Gamma_{\text{ref},n}^{\text{U}},\delta_{\text{vir}}^{\text{L}}\right),
\end{eqnarray}
where $\Gamma^{\text{U}}_n$ ($\Gamma^{\text{U}}_{\text{ref},n}$) is an upper bound on the phase error probability of the actual (reference) states in the round $n$. Then, by using Jensen's inequality~\cite{jensen1906fonctions}, we obtain
\begin{eqnarray}\label{eq:Ap:Gamman1}
\frac{1}{N}\sum_n\Gamma^{(n)}_{\mathcal{K}}&=&\frac{1}{N}\sum_np_{\mathcal{K}} G_+\left(\Gamma_{\text{ref},n}^{\text{U}},\delta_{\text{vir}}^{\text{L}}\right)\nonumber\\
&\leq& p_{\mathcal{K}} G_+\left(\frac{1}{N}\sum_n\Gamma_{\text{ref},n}^{\text{U}},\delta_{\text{vir}}^{\text{L}}\right),
\end{eqnarray}
due to the concavity of $G_+$ with respect to its first element. Now, we can take advantage of the fact that the function $\Gamma_{\text{ref}}^{\text{U}}$ given in Eq.~(\ref{eq:GammaU}) is also concave with respect to $Y_{\nu,\omega}$, which for a particular round $n$ we denote as $Y_{\nu,\omega}^{n}$, and apply again Jensen's inequality, now to $1/N\sum_n\Gamma_{\text{ref},n}^{\text{U}}$, so we have
\begin{widetext}
\begin{eqnarray}\label{eq:Ap:Gamman2}
\frac{1}{N}\sum_n\Gamma_{\text{ref},n}^{\text{U}}&=&\frac{1}{N}\sum_n\sum_{\substack{\nu,\omega \\ f_{\nu,\omega}>0}}f_{\nu,\omega}G_ +\left(\frac{\tilde{Y}_{\nu,\omega,\mathcal{T}}^n}{p_{\nu,\omega}p_{\mathcal{T}|\nu,\omega}},\delta_{\nu,\omega}^{\rm L}\right) +\frac{1}{N}\sum_n\sum_{\substack{\nu,\omega \\ f_{\nu,\omega}<0}}f_{\nu,\omega}G_ -\left(\frac{\tilde{Y}_{\nu,\omega,\mathcal{T}}^n}{p_{\nu,\omega}p_{\mathcal{T}|\nu,\omega}},\delta_{\nu,\omega}^{\rm L}\right)\nonumber\\
&\leq&\sum_{\substack{\nu,\omega \\ f_{\nu,\omega}>0}}f_{\nu,\omega}G_ +\left(\frac{\sum_n\tilde{Y}_{\nu,\omega,\mathcal{T}}^n}{Np_{\nu,\omega}p_{\mathcal{T}|\nu,\omega}},\delta_{\nu,\omega}^{\rm L}\right)+\sum_{\substack{\nu,\omega \\ f_{\nu,\omega}<0}}f_{\nu,\omega}G_ -\left(\frac{\sum_n\tilde{Y}_{\nu,\omega,\mathcal{T}}^n}{Np_{\nu,\omega}p_{\mathcal{T}|\nu,\omega}},\delta_{\nu,\omega}^{\rm L}\right),
\end{eqnarray}
\end{widetext}
where $\tilde{Y}_{\nu,\omega,\mathcal{T}}^n:=Y_{\nu,\omega}^{n}p_{\nu,\omega}p_{\mathcal{T}|\nu,\omega}$ is the joint probability that Alice and Bob send $\ket{\Psi_{\nu,\omega}}$, Charles announces a successful event in the round $n$, and the round is used for parameter estimation, $p_{\nu,\omega}=p_{\nu}p_{\omega}$, and $p_{\mathcal{T}|\nu,\omega}$ is the conditional probability that the round is used for parameter estimation given that Alice and Bob send $\ket{\nu}$ and $\ket{\omega}$, respectively. Note that $p_{\mathcal{T}|\nu,\omega}=1$ if any of Alice or Bob's states is the vacuum state. Also, we have that $p_{\mathcal{K}}=1-\sum_{\nu,\omega}p_{\nu,\omega}p_{\mathcal{T}|\nu,\omega}$. By combining Eqs.(\ref{eq:Ap:Gamman1}) and~(\ref{eq:Ap:Gamman2}), one arrives to the following bound
\begin{widetext}
\begin{equation}\label{eq:Ap:Gamma3}
	\sum_n\Gamma^{(n)}_{\mathcal{K}}\leq Np_{\mathcal{K}} G_+\left(\sum_{\substack{\nu,\omega \\ f_{\nu,\omega}>0}}f_{\nu,\omega}G_ +\left(\frac{\sum_n\tilde{Y}_{\nu,\omega,\mathcal{T}}^n}{Np_{\nu,\omega}p_{\mathcal{T}|\nu,\omega}},\delta_{\nu,\omega}^{\rm L}\right) +\sum_{\substack{\nu,\omega \\ f_{\nu,\omega}<0}}f_{\nu,\omega}G_ -\left(\frac{\sum_n\tilde{Y}_{\nu,\omega,\mathcal{T}}^n}{Np_{\nu,\omega}p_{\mathcal{T}|\nu,\omega}},\delta_{\nu,\omega}^{\rm L}\right),\delta_{\text{vir}}^{\text{L}}\right).
\end{equation}
\end{widetext}
Importantly, the probability $\tilde{Y}_{\nu,\omega,\mathcal{T}}^n$ could depend on all the available information up to the $n$-th round. This means that, with a negligible probability of failure for $N\to\infty$, one can estimate the sums $\sum_n\tilde{Y}_{\nu,\omega,\mathcal{T}}^n$ from the observed number of successful events within the parameter estimation rounds where Alice and Bob send $\ket{\Psi_{\nu,\omega}}$, namely $\tilde{N}_{\nu,\omega,\mathcal{T}}$, by using Azuma's inequality~\cite{azuma1967weighted} or Kato's inequality~\cite{kato2020concentration}. Specifically,  $\tilde{N}_{\nu,\omega,\mathcal{T}}\approx \sum_n\tilde{Y}_{\nu,\omega,\mathcal{T}}^n$.
\\
\\
Finally, it is possible to obtain an estimation on the number of phase errors, $\tilde{N}_{\text{ph}}$, from the sum $\sum_n\Gamma^{(n)}_{\mathcal{K}}$ by applying again Azuma's or Kato's inequality. That is, we have that $\tilde{N}_{\text{ph}}\approx \sum_n\Gamma^{(n)}_{\mathcal{K}}$ with negligible probability of failure when $N\to\infty$.

\section{Optimal amplitude $\alpha$}\label{appendix:alpha}
\begin{figure}[h!]
	\begin{center}
		\includegraphics[width=\columnwidth]{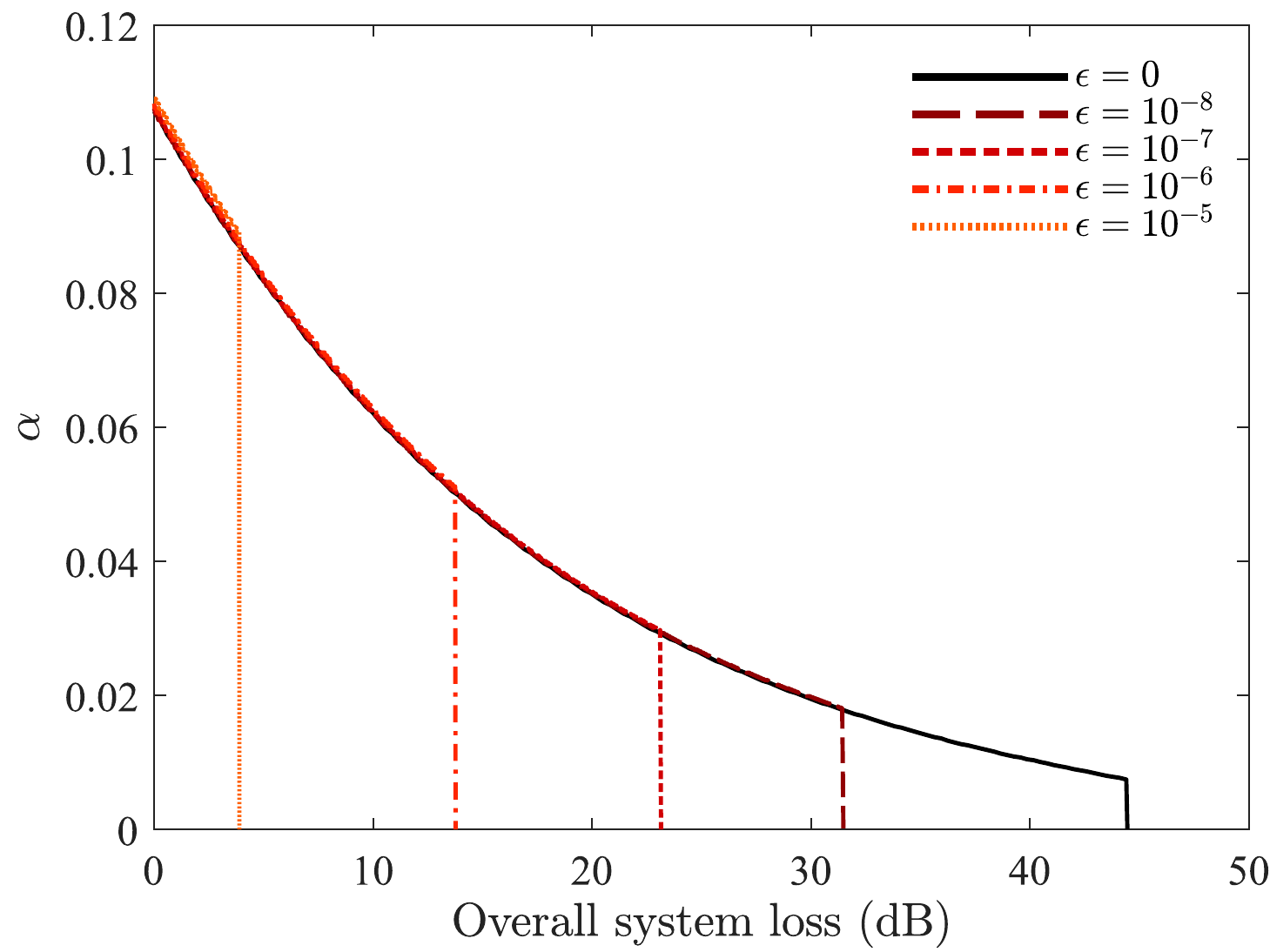}
	\end{center}
	\caption{Optimal value of $\alpha$ corresponding to the simulations shown in Fig.~\ref{fig:Fig_Results} in the main text.} \label{fig:Fig_Optimal_alpha} 
\end{figure}

Here we show in Fig.~\ref{fig:Fig_Optimal_alpha}, for completeness, the optimized values of the parameter $\alpha$ corresponding to the simulations shown in Fig.~\ref{fig:Fig_Results} of the main text.

\section{Non-vacuum intensity}\label{appendix:vacuum}
\begin{figure}
	\begin{center}
		\includegraphics[width=\columnwidth]{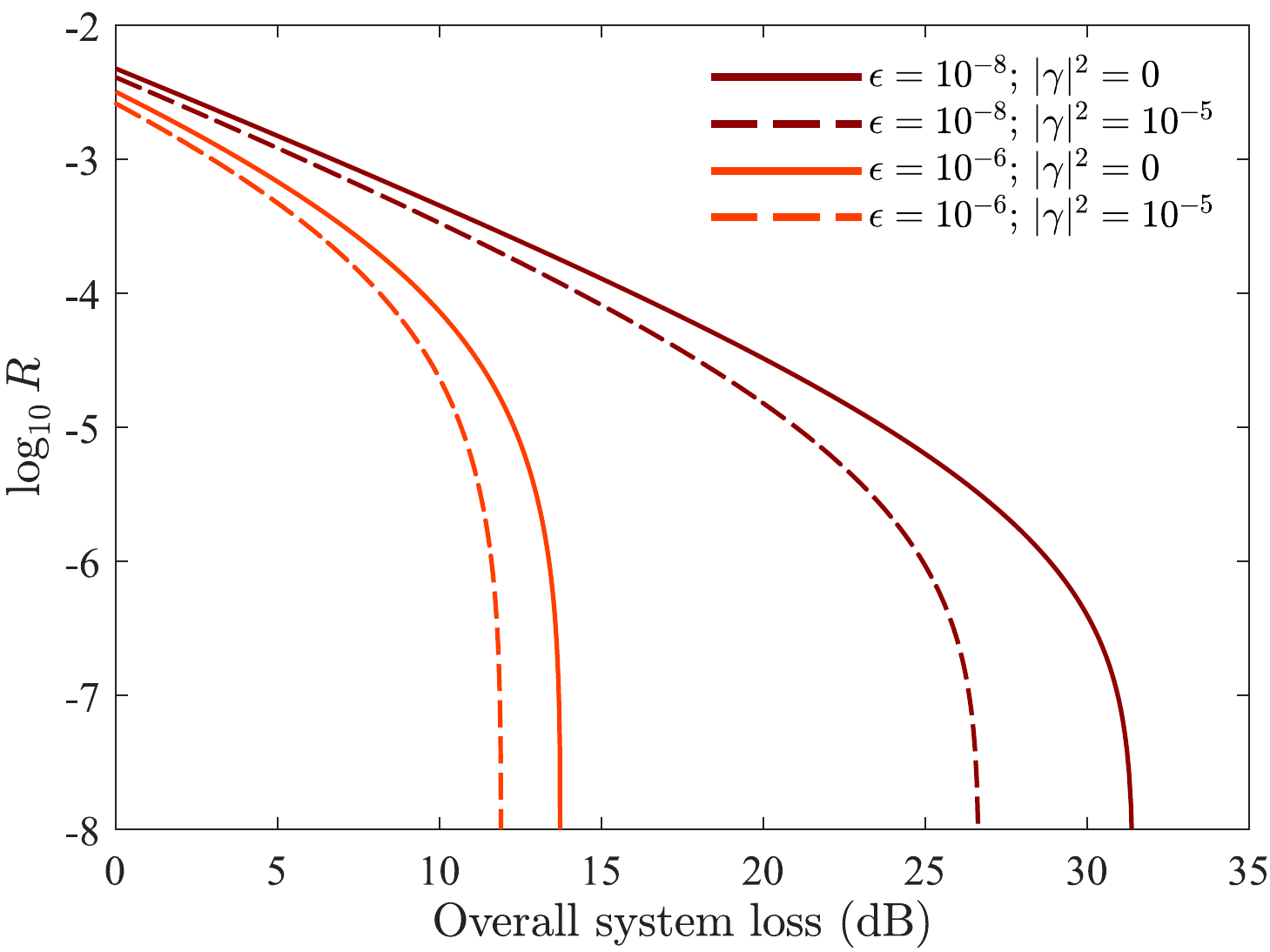}
	\end{center}
	\caption{Comparison between the ideal scenario where the third reference state $\ket{\gamma}$ used by Alice and Bob is a perfect vacuum state ($\abs{\gamma}^2=0$) and the case where, instead, such state is a weak coherent state ($\abs{\gamma}^2=10^{-5}$). As it can be observed, the performance of the protocol is similar in both cases.} \label{fig:Fig_Comparison_Intensity}
\end{figure}

Here we illustrate the effect that the use of imperfect vacuum states has on the performance of the protocol. For this, we consider the secret key rate that Alice and Bob would obtain when they use the set of states $\set{\ket{\alpha},\ket{-\alpha},\ket{\gamma}}$, with $\gamma\in\mathbb{R}$. That is, this set of states is used in the simulations to calculate the experimental probabilities $Y_{\nu,\omega}$ as well as to define the set of reference states $\set{\ket{\alpha},\ket{-\alpha},\ket{\gamma'}}$, being $\ket{\gamma'}$ the projection of $\ket{\gamma}$ onto the qubit space spanned by $\set{\ket{\alpha},\ket{-\alpha}}$. Note, however, that any $\abs{\gamma}^2>0$ could also be treated as an imperfection and thus it could be incorporated to the security proof by properly choosing the parameters $\epsilon_{\nu,\omega}$. The parameter $\alpha$ is optimized for each value of the overall system loss, and for illustration purposes we evaluate two cases for the intensity $\abs{\gamma}^2$: 0 and $10^{-5}$. As one can see in Fig.~\ref{fig:Fig_Comparison_Intensity}, the performance is very similar in both cases, only slightly lower when $\abs{\gamma}^2=10^{-5}$.

\end{document}